\documentclass[%
 aip,
 amsmath,amssymb,
 reprint,%
]{revtex4-2}
\usepackage{graphicx,tabularx}% Include figure files
\usepackage{dcolumn}% Align table columns on decimal point
\usepackage{bm}% bold math
\usepackage{graphicx,subcaption,}% Include figure files
\usepackage{dcolumn}% Align table columns on decimal point
\usepackage{bm}% bold math
\usepackage{color}

\usepackage[utf8]{inputenc}
\usepackage[T1]{fontenc}
\usepackage{mathptmx}
\usepackage{etoolbox}

\makeatletter
\def\@email#1#2{%
 \endgroup
 \patchcmd{\titleblock@produce}
  {\frontmatter@RRAPformat}
  {\frontmatter@RRAPformat{\produce@RRAP{*#1\href{mailto:#2}{#2}}}\frontmatter@RRAPformat}
  {}{}
}%
\makeatother
\begin{document}

\preprint{AIP/123-QED}

%\title[Sample title]{Sample Title:\\with Forced Linebreak}
%\title[Sample title]{Estimates of the evolution of a longitudinal bore in PMMA bars...?}
\title{Theoretical estimates of the parameters of longitudinal undular bores in PMMA bars based on their measured initial speeds}

% Force line breaks with \\
\author{Curtis G. Hooper}
\affiliation{ 
	Wolfson School of Mechanical, Electrical and Manufacturing Engineering, Loughborough University, Loughborough, LE11 3TU, United Kingdom%\\This line break forced with \textbackslash\textbackslash
}%Lines break automatically or can be forced with \\
\affiliation{ 
	Department of Mathematical Sciences, Loughborough University, Loughborough, LE11 3TU, United Kingdom%\\This line break forced with \textbackslash\textbackslash
}%
\author{Karima R. Khusnutdinova}%
\affiliation{ 
Department of Mathematical Sciences, Loughborough University, Loughborough, LE11 3TU, United Kingdom%\\This line break forced with \textbackslash\textbackslash
}%

\author{Jonathan M. Huntley}
% \homepage{http://www.Second.institution.edu/~Charlie.Author.}
\affiliation{ 
	Wolfson School of Mechanical, Electrical and Manufacturing Engineering, Loughborough University, Loughborough, LE11 3TU, United Kingdom%\\This line break forced with \textbackslash\textbackslash
}%

\author{Pablo D. Ruiz$^*$ }
%\homepage{http://www.Second.institution.edu/~Charlie.Author.}
\affiliation{ 
	Wolfson School of Mechanical, Electrical and Manufacturing Engineering, Loughborough University, Loughborough, LE11 3TU, United Kingdom%\\This line break forced with \textbackslash\textbackslash
}%
 \email{P.D.Ruiz@lboro.ac.uk}

\date{\today}% It is always \today, today,

\begin{abstract}
We study the evolution of the longitudinal release wave that is generated by induced tensile fracture as it propagates through solid rectangular Polymethylmethacrylate (PMMA) bars of different constant cross section. High speed multi-point photoelasticity is used to  register the strain wave.  In all cases, oscillations develop at the bottom of the release wave that exhibit the qualitative features of an undular bore. The pre-strain, post-strain, strain rate of the release wave and %width of the cross section of the bar contribute to the evolution of the oscillations. %The leading oscillation can be described by simple formulae derived from the analytical solution of the linear Korteweg - de Vries equation.
 the cross section dimensions determine the evolution of the oscillations.
%On finding 
From the wave speed and strain rate close to the fracture site, %from experimental data, 
we estimate %some features 
the strain rate of the release wave as well as the growth of the amplitude and duration of the leading %part of the bore
oscillation away from the fracture site on using formulae derived from the  simple analytical solution of the  linearised Gardner equation  (linearised near the pre-strain level at fracture), developed in our earlier work \cite{HRHK}. Our estimates are then compared to experimental data, where qualitative and good semi-quantitative agreements are established.
\end{abstract}

\maketitle

\section{\label{sec:level1}Introduction}
Recently, we reported our first observations and modelling of longitudinal undular bores in PMMA bars generated by natural and induced tensile fracture, where observations at different distances from the fracture site were obtained in individual experiments using the single-point photoelasticity. \cite{HRHK} Here, we significantly improve the experimental methodology by developing the high speed multi-point photoelasticity, and use this methodology to study the dependence of the key parameters of the undular bores 
%the undular bores 
generated by the induced tensile fracture of PMMA bars on the dimensions of rectangular cross sections.

Undular bores (or dispersive shock waves) are non-stationary waves which propagate as an oscillatory transition between two basic states. The key characteristics of the oscillatory structure are that it gently expands and grows in amplitude with propagation distance. They have been  studied both experimentally and theoretically in  various physical contexts (see, for example, \cite{WJF,K_et_al, EH,TKCO,B_et_al,VMS} and references therein).
Undular bores form naturally in nature and have been photographed and identified in rivers around the world \cite{Chanson_2011} as well as in the atmosphere through cloud formations \cite{Coleman_2010}. Perhaps the most striking of these phenomena are the famous \emph{Morning Glory} cloud formations which  are frequently observed off the northern coast of Australia \cite{Reid_1981}. 

 Similar wave structures have previously been observed in solids during a variety of impact test experiments, but  were only recently identified and described as undular bores, as a by-product of our work on undular bores generated by fracture \cite{HRHK}.  The clearest examples were observed in long metal waveguides, where an oscillatory transition region develops between the basic state of rest and the basic state of compressive strain due to geometrical dispersion in the waveguide \cite{Kolsky_1958,Ren_army, SJ}. The qualitative features of an undular bore (expansion and growing amplitude of the oscillatory structure close to the transition) are observed.
%which are not viscoelastic 
Similar waves have also been registered and studied in PMMA \cite{ZGK,Wang}.
% It was only recently that undular bores were officially reported in solids after they were generated by tensile fracture in rectangular PMMA bars of constant cross section \cite{HRHK_2021}.

In this paper, we extend the tensile fracture experiments 
%in our earlier work \cite{HRHK} 
by observing the waves generated by the tensile fracture of pre-strained PMMA bars of different rectangular cross sections using multi-point high speed photoelasticity and elaborate on features of the bore within the scope of the simple solution of the ``small on large" model obtained by linearising the nonlinear model equation near the level of the pre-strain at fracture. 
%linear elasticity. 
Estimates of the growth of amplitude and duration of the leading oscillation are given %on
 using the wave speed, strain rate, pre-strain and post-strain close to the fracture site,  and compared with the experimental data.

\section{Experiment}
\begin{figure}[t]
	%\centering
	\begin{center}
		\includegraphics[width=8.1cm]{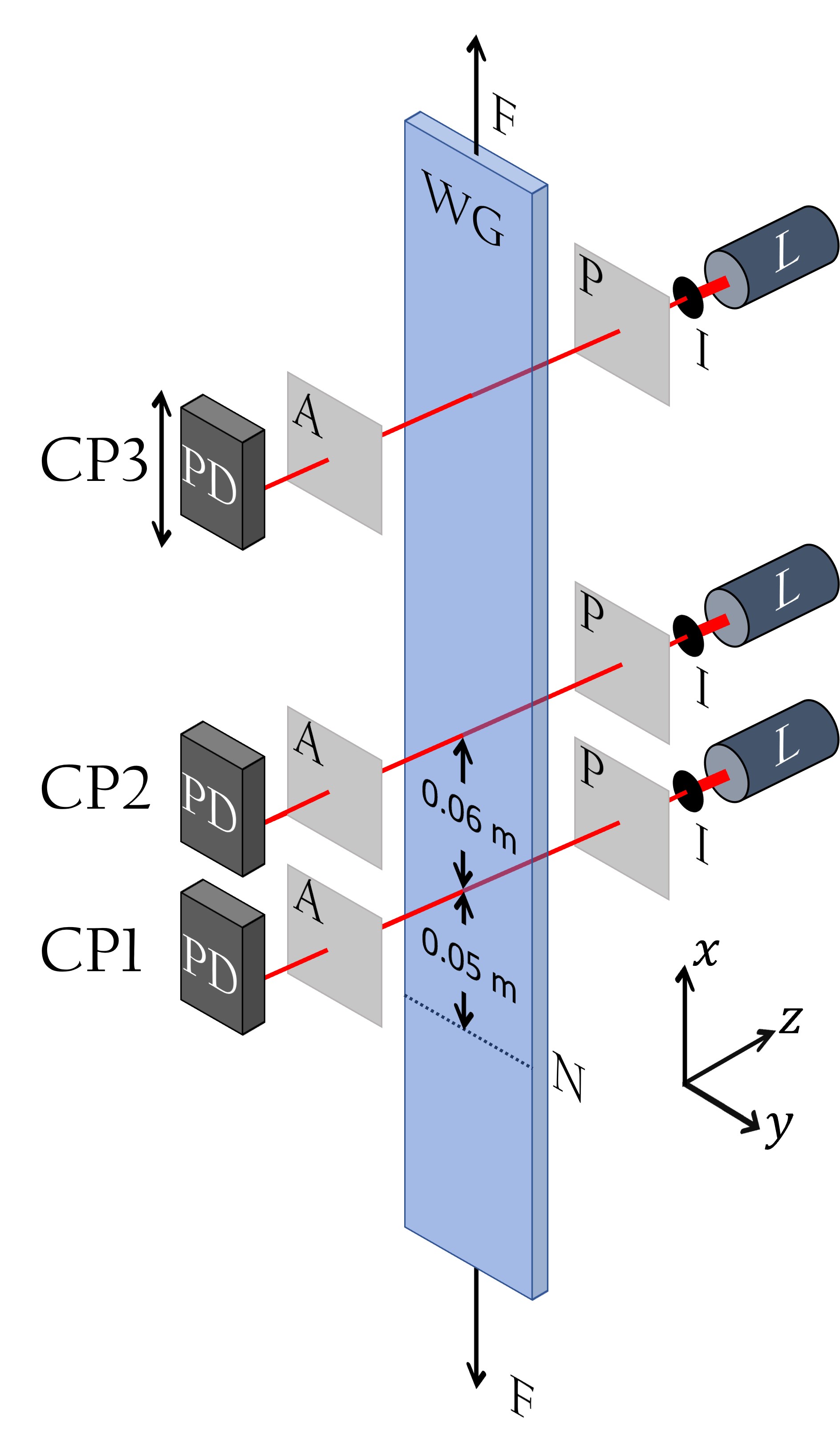}
	\end{center}
	\caption{Experimental setup, showing the PMMA waveguide (WG) under tensile load (F) with a light notch (N) at the fracture site at $x=0$, and circular polariscopes (CP1, CP2, CP3) each consisting of a laser diode (L), iris (I), polarizer (P), analyser (A) and photodetector (PD).}
	\label{fig:exp}
\end{figure}

%Rectangular 
PMMA bars of uniform rectangular cross section were cut from the same sheet of material. Three different cross sections were cut, resulting in bars of dimension
3 $\times$ 10 $\times$ 770 mm$^3$, 3 $\times$ 15 $\times$ 770 mm$^3$  and 3 $\times$ 20 $\times$ 770 mm$^3$. The bars were loaded at a constant strain rate of $1 \times 10^{-3}$ s$^{-1}$ until a stress of 50 MPa was applied using a tensile testing machine (TTM, Instron 3345). %with a 5 kN load cell. 
The corresponding strain in all bars at this level of stress was $\sim 0.02$. The stress was then held at 50 MPa until fracture was induced by pressing a blade against the sample, 0.05 m above the lower grip, and running it across a very shallow pre-notched groove in the 10 mm, 15 mm or 20 mm width of the bar. The notch was sufficiently shallow to not cause natural tensile fracture of the bar during loading%. It serves 
, but served merely as a guide for the blade. Once loaded into the TTM, the length of the sample between the grips was 720 mm.
A schematic of the experimental arrangement can be seen in Fig. \ref{fig:exp}.%, in a horizontal configuration. 

					\begin{figure}[t!]
	\centering
	\begin{subfigure}[t!]{0.5\textwidth}
		\includegraphics[width=0.95\textwidth]{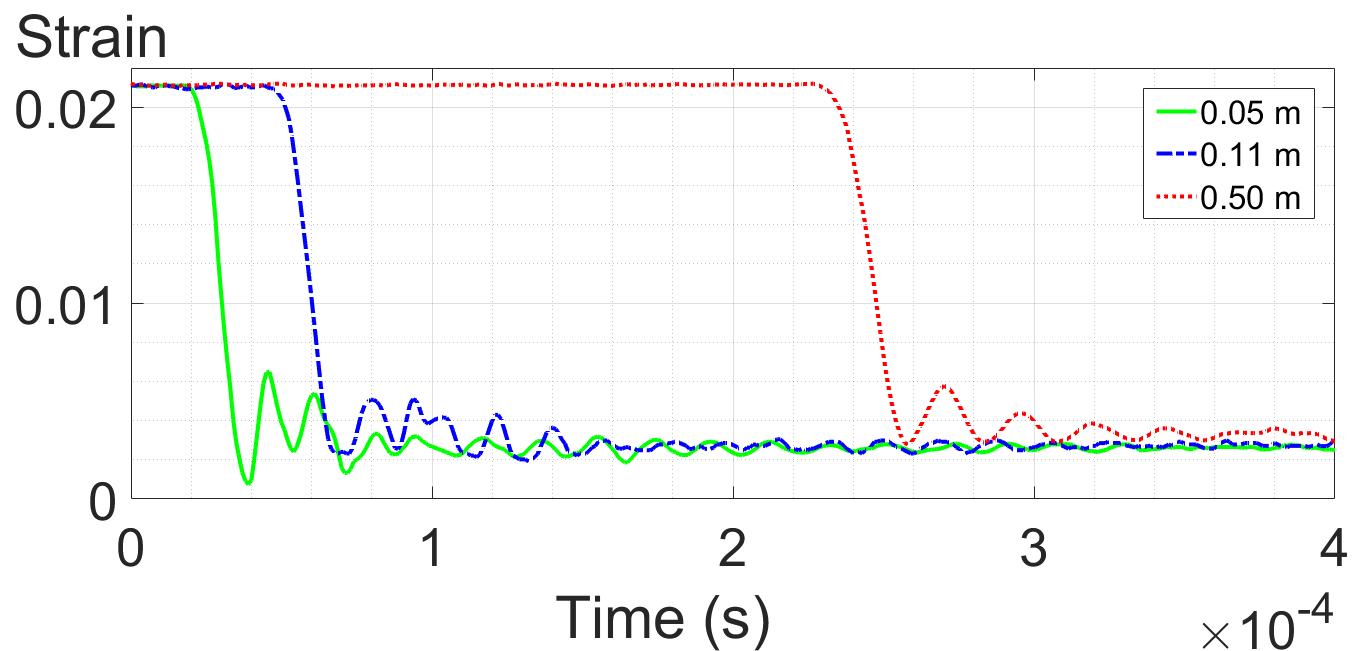}
		\caption{%Raw strain profiles.
		}
		\label{fig:2a}
	\end{subfigure}	
	
	\begin{subfigure}[t!]{0.5\textwidth}
		\centering
		\includegraphics[width=0.95\textwidth]{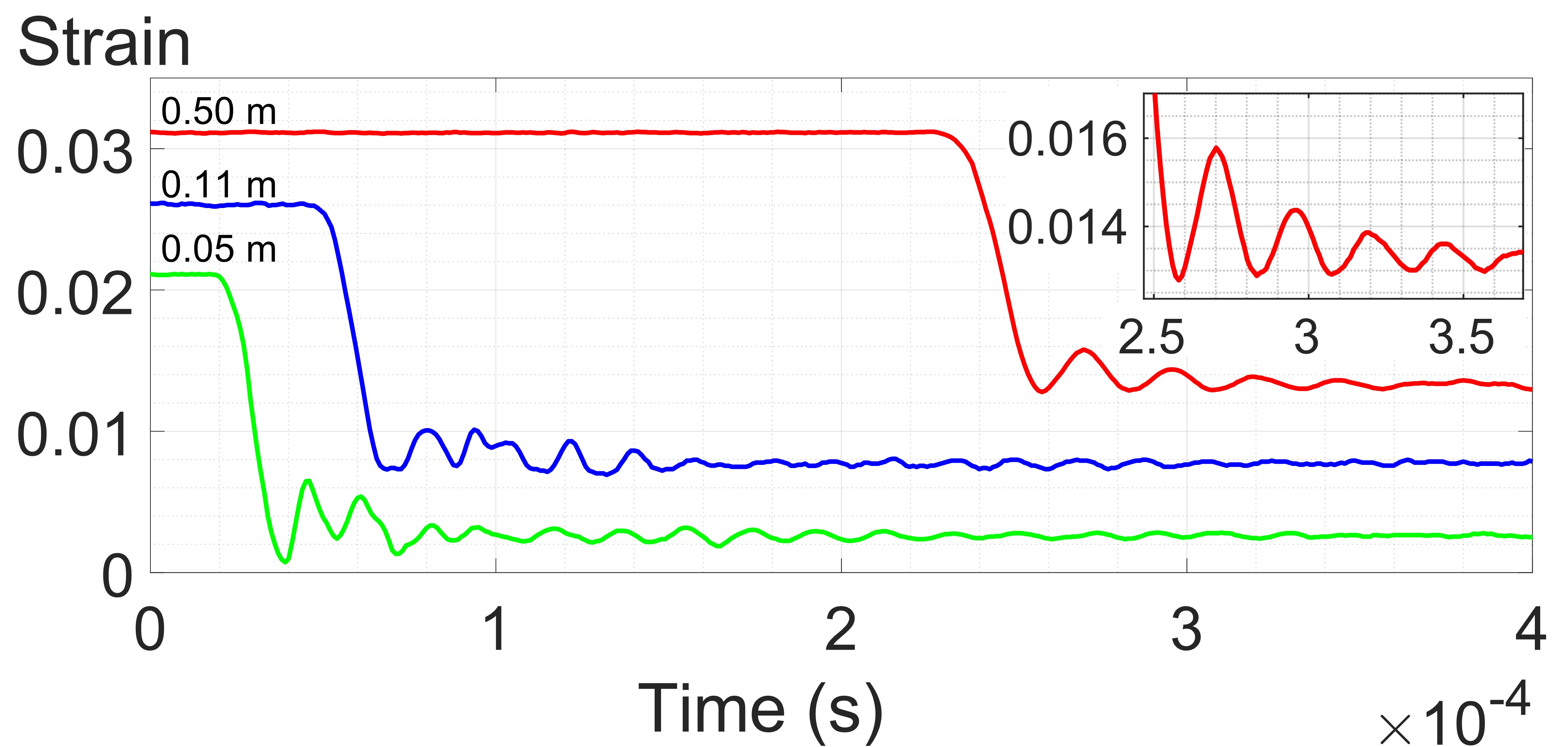}
		\caption{%Raw strain profiles with equal vertical spacing for separation. The insert shows the developing oscillations at 0.50 m.
		}
		\label{fig:2b}
	\end{subfigure}% \ \
	\caption{Experimental strain profiles after fracture of a 20 mm wide bar at distances of 0.05 m, 0.11 m and 0.50 m from the fracture site, recorded by CP1, CP2 and CP3 respectively. All profiles are recorded from the same experiment. Time $t = 0$ s corresponds to the time of fracture. A vertical offset of 0.005 has been added in (b) for clarity.}
	\label{fig:2}
\end{figure}

%PMMA is birefringent under strain, thus photoelasticity \cite{photoelasticity1} was used to measure strains  via a polariscope.  We mounted a circular polariscope in bright field configuration (CP) onto the end of a rail which was clamped onto a vertical column that was secured next to the TTM. The CP could be secured to different positions on the vertical column allowing the intensity at different distances from fracture to be recorded. The light source was a 632 nm 30 mW He-Ne laser.  A polarisation maintaining optical fiber with an angled physical connection was used to couple the beam from the laser aperture to the CP. This fiber ensured the intensity of light incident on the bar was constant and not altered by minor vibrations or temperature changes around the apparatus. This was crucial as we required changes in intensity to be caused only by changes in strain in the bar. The laser beam was collimated to a diameter of 1 mm before continuing through a polariser. It then passed through the bar and an analyser before being captured by a Thorlabs Si switchable gain photodetector (PD). The voltage, which is directly proportional to intensity, was displayed on a laptop via a picoscope 2000 digital oscilloscope.  The gain on the PD was set to 0 dB, and a 50 $\Omega$ terminating resistor was connected which enabled the PD to achieve its maximum bandwidth of 12 MHz. A simple triggering circuit was made using a power supply and a 1 k$\Omega$ resistor so the time of fracture was known. 

As PMMA exhibits transient birefringence \cite{photoelasticity1}, three bright-field circular polariscopes (CP1, CP2, CP3) were used to measure the longitudinal strain in the bar. Each CP consisted of a %usb powered 
laser %module light 
source (Thorlabs PL202, 635 nm, 0.9 mW), an iris to reduce the beam diameter to 2 mm, a circular polarizer (P), a circular analyser (A) and a photodetector (PD,  Thorlabs PDA36A2, 350 - 1100 nm, 12 MHz bandwidth). The distance between the laser beam of CP1 and CP2 was a fixed %0.06 m 
60 mm. They were positioned %0.05 m 
50 mm and %0.11 m 
110 mm above the fracture site. CP3 was mounted on a platform that could be moved and secured at different distances from the fracture site (indicated by the double arrow in Fig. \ref{fig:exp}).
% so that the strain transient due to fracture can be measured at different distances from the notch. 

%The time of fracture was determined by calculating the speed of the front of the release wave from the three recorded intensity profiles.

The setup allowed us to observe the initial evolution of the wave close to the fracture site with CP1 and CP2, and also the long time evolution of the wave with CP3 which was positioned further away from the fracture site. 
 
\subsection{Strain evaluation}
%Now an equation linking intensity and strain must be derived. 
In our setting, the light intensity at the photodetector is given by
\begin{equation}
	I = I_0\cos^2\left(\frac{\pi h (\sigma_x - \sigma_y)}{f_\sigma} \right ),
	\label{eqn:id}
\end{equation}
where $I_0$ is the intensity of the laser corresponding to zero stress, $h = 3$ mm is the sample thickness of the bar in the direction of the laser beam, $\sigma_{x, y}$ are the respective values of stress and $f_\sigma$ is the fringe constant of the material \cite{photoelasticity}.
%Under uniaxial stress loading,  $\sigma_y = 0$, and the  longitudinal strain $e_x$ and stress $\sigma_x$ are related by Hooke's law 
Under uniaxial loading and for the relevant timescales and sample widths in this study, $\sigma_y$ is effectively zero, and the  longitudinal strain $e_x$ and stress $\sigma_x$ are related by Hooke's law 
%via the formulae
%\noindent
%\begin{eqnarray*}
%${\displaystyle \sigma_x = \frac{E}{1-\nu^2}(\epsilon_x+\nu\epsilon_y), \hspace{0.1cm} 
%\label{eqn:sigma1} \quad 
%	\sigma_y = \frac{E}{1-\nu^2}(\epsilon_y+\nu\epsilon_x),}$
%\label{eqn:sigma2}
%\end{eqnarray*}
%	consistent with the assumptions that waves are long compared to $h$ and weakly nonlinear, made below.
%	Here, $E$ and $\nu$ are Young's modulus and Poisson's ratio of the PMMA, $\sigma_x$ and $\sigma_y$ are the stresses in the longitudinal and transverse directions, whilst $\epsilon_x$ and $\epsilon_y$ are the respective strains. In our experiment, $\sigma_y=0$, which yields $\epsilon_y = -\nu\epsilon_x$ and 
$ \sigma_x = E e_x$, %consistent with the assumptions that waves are long compared to $h$ and weakly nonlinear, made below. 
where $E$ is Young's modulus. %Thus equation \eqref{eqn:id} becomes
%Thus,
%\begin{equation}
%${\displaystyle I = I_0 \cos^2 \Big(\frac{\pi h E e_x}{f_\sigma}\Big ).}$}
%which is used to reconstruct longitudinal strain $\epsilon_x$ as a function of intensity. 
When equation (\ref{eqn:id}) is rearranged for $e_x$, we have
\begin{equation}
	e_x = \frac{f_\sigma}{\pi E h} \left[ \cos^{-1}\left(\pm \sqrt{\frac{I(t)}{I_0}}\right) + N\pi \right],
	\label{eqn:strain}
\end{equation}
where $N$ is the (integer) fringe order. Two branches are introduced in the form of the $\pm$ in the argument of the $\displaystyle{\cos^{-1}}$ function, and the addition of an integer multiple of $\pi$. For correct construction of the strain during the loading period, the $'+'$ sign is chosen when the intensity is decreasing and the $'-'$ sign is applied when the intensity is increasing. The value $N$ is initially 0 and increases by 1 whenever a fringe is completed. After fracture, $N$ decreases by 1 on completion of a fringe, $'-'$ corresponds to decreasing intensity and $'+'$ corresponds to increasing intensity.

\begin{table}
	\caption{\label{Tab_1}The values of pre-strain ($\kappa$), post-strain at 0.05 m ($\kappa_{t_{5}}$), strain rate at 0.05 m ($\dot{e}_{5}$), strain rate at 0.50 m ($\dot{e}_{50}$), and the average speed of the top of the release wave ($0.95\kappa$) between 0.05 m and 0.11 m ($v_1$)  for the strain profiles shown in Fig. \ref{fig:2a}.}
	\begin{ruledtabular}
	\begin{tabular}{cccccc}
	$\kappa$&$\kappa_{t_{5}}$&$\kappa_{t_{50}}$&$\dot{e}_{5}$ (s$^{-1}$)&$\dot{e}_{50}$ (s$^{-1}$)&$v_{1}$ (m/s)\\
	\hline
	0.021 &0.0026 &0.0033 &1640&1090&2120\\
\end{tabular}
	\end{ruledtabular}
\end{table}
% v2 = 2110 m/s

\section{Results}
Over a series of tests, CP3 was placed between 0.20 m and 0.50 m from the fracture site.
Before the intensity can be converted to strain through equation \eqref{eqn:strain}, the values of $E$ and $f_\sigma$ must be calculated. We will denote the value of the quasi-static Young modulus by $E_0$, and use $E$ to denote the dynamic Young modulus, which is dependent on the strain rate (see our earlier paper \cite{HRHK} and the references therein).

To %calculate 
determine $E_0$ prior to fracture we looked to the stress-strain curve produced during the loading period where the stress increased from 0 Pa to 50 MPa. At the strain rate of $1 \times 10^{-3}$ s$^{-1}$, the stress / strain curve for PMMA is approximately linear up to strains of $\sim 0.02$ \hspace{0.05cm} \cite{WMX}. %So we fitted a linear line of best fit to the curve using the data fitting toolbox in MATLAB, and took the gradient of the line as the value of $E$. 
The value of $E_0$ was established from the slope of a linear fit to the measured stress / strain curve. From the entire set of results, the average value of $E_0$ was $2.4 \pm 0.076$  GPa. 

On the completion of 1 fringe when the stress in the $x$ direction is equal to $\sigma_x^1 \neq 0$, the intensity resumes the value of $I_0$, and from \eqref{eqn:id} we have that
\begin{equation}
	\frac{\pi h \sigma_x^1}{f_\sigma} = \pi,
\end{equation}
so $f_\sigma = h \sigma_x^1$. The average from the entire data set was $f_\sigma = 1.2 \times 10^5$ Pa$\times$m/fringe.

After fracture when the strain rate dramatically increases, it is known that the value of $E$ will also increase \cite{WMX, Li_Lambros, Ahzi_06}. To ensure the continuity of both \eqref{eqn:id} and \eqref{eqn:strain}, $f_\sigma$ will increase too such that the ratio of $f_\sigma / E$ is constant throughout to maintain the continuity of strain. Therefore the values of $E$ and $f_\sigma$ obtained here from the loading period are used to convert the intensity observed after fracture to strain. An example of the strain profiles obtained from a typical test after fracture are shown in Fig. \ref{fig:2}. %A typical result is shown in Figure \ref{fig:2}. %Details of 3 tests are shown in Table \ref{Tab_1}.

After fracture, a longitudinal release wave propagates radially outwards from the fracture site \cite{Fabrice_2020}. At each distance, no relaxation is observed at times before the arrival of the release wave.  When the release wave arrives at a polariscope, the strain decreases from the pre-strain level to a temporary non-zero level which we refer to as the post-strain.  %For the times relevant to our experiments, the post-strain is constant at each distance, although it does appear to increase with distance from the fracture site (see right of Fig. \ref{fig:2a}). Thus the strain decreases by a greater amount closer to fracture than it does further away from fracture in this time period.
  It is calculated as the average strain from a $1\times 10^{-4}$ s window sufficiently far away from any oscillations. %Total relaxation is eventually observed (in the order of seconds), thus there are no signs of plasticity.  
  In what follows, the pre- and post-strain levels are denoted by $\kappa$ and $\kappa_t$ respectively.

The duration of the initial unloading, and therefore the strain rate of the release wave close to the fracture site, are determined by the distance the crack tip propagates to cause fracture, and the speed at which it propagates. %Crack propagation in viscoelastic materials has been extensively studied \cite{Willis, MW, Fleck_2018}. Once initiated, cracks accelerate very rapidly to propagation speeds between $150$ m s$^{-1}$ and $300$ m s$^{-1}$ in PMMA\cite{Takahashi_1984, Doll_1976, Sheng_1999, Wang_2019}. 
We look to the lower part of the release wave to calculate the strain rate as this is where the longitudinal oscillations develop from, %that 
which are discussed later. Specifically we calculate it from the region given by $0.25(\kappa + \kappa_t)$ and $0.5(\kappa + \kappa_t)$. We find the strain rate %is less 
decreases at distances further away from the fracture site, thus the slope gets shallower as the wave propagates.

%\begin{figure}
%	\centering
%	\includegraphics[width=0.5\textwidth]{Fig_3}
%	\caption{Two experimental strain profiles with different strain-rates and similar pre-strains from 15 mm wide bars, both recorded by CP3 when it was positioned 0.30 m away from the fracture site. In the insert, the red (solid) curve has been translated in order to provide an easier comparison of the oscillations. }
%	\label{fig:fig3}
%\end{figure}

					\begin{figure}[t!]
	\centering
	\begin{subfigure}[t!]{0.5\textwidth}
		\includegraphics[width=0.95\textwidth]{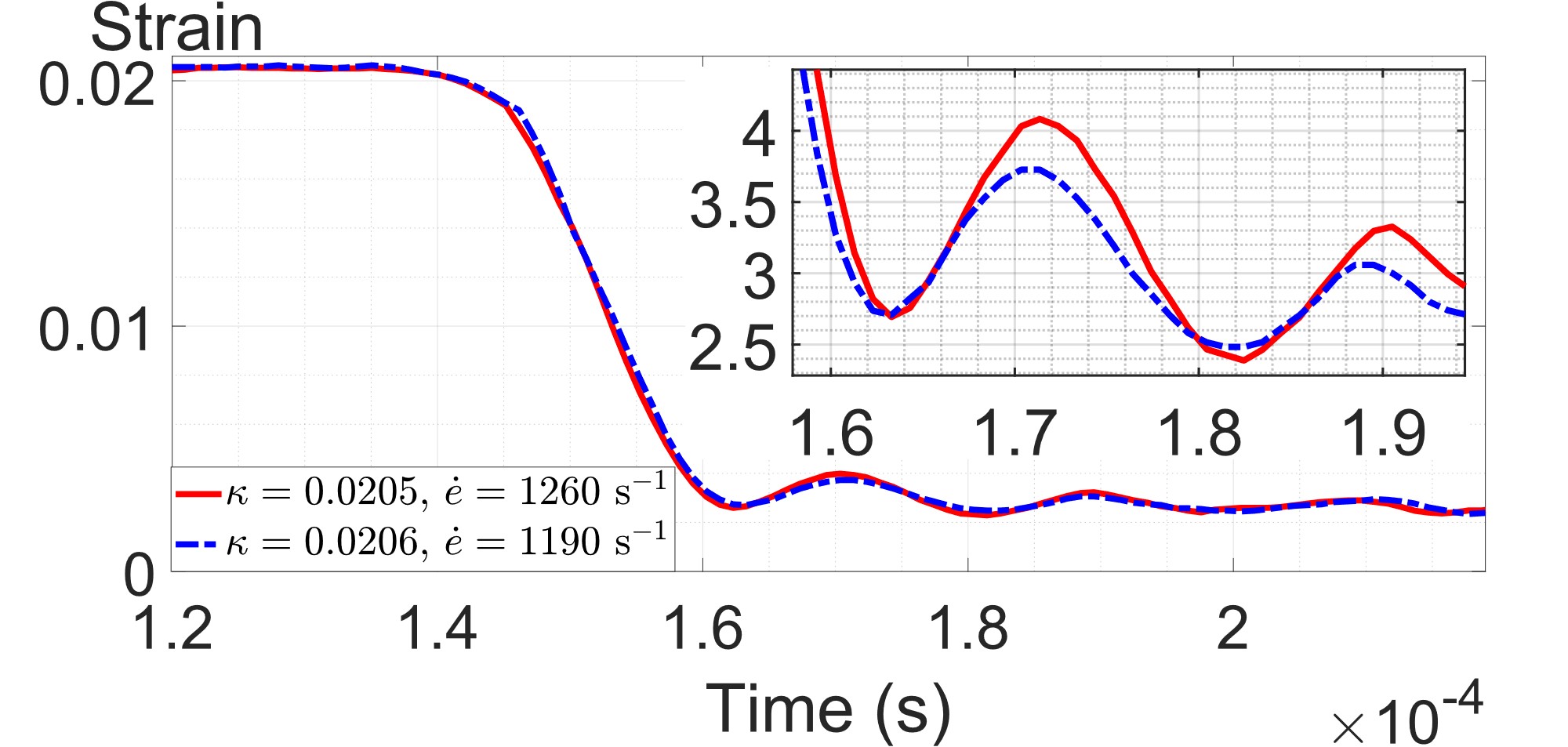}
		\caption{Two experimental strain profiles with different strain-rates and similar pre-strains from 15 mm wide bars, both recorded by CP3 when it was positioned 0.30 m away from the fracture site. In the insert, the red (solid) curve has been translated in order to provide an easier comparison of the oscillations.}
		\label{fig:fig3}
	\end{subfigure}	

	\begin{subfigure}[t!]{0.5\textwidth}
	\centering
	\includegraphics[width=0.95\textwidth]{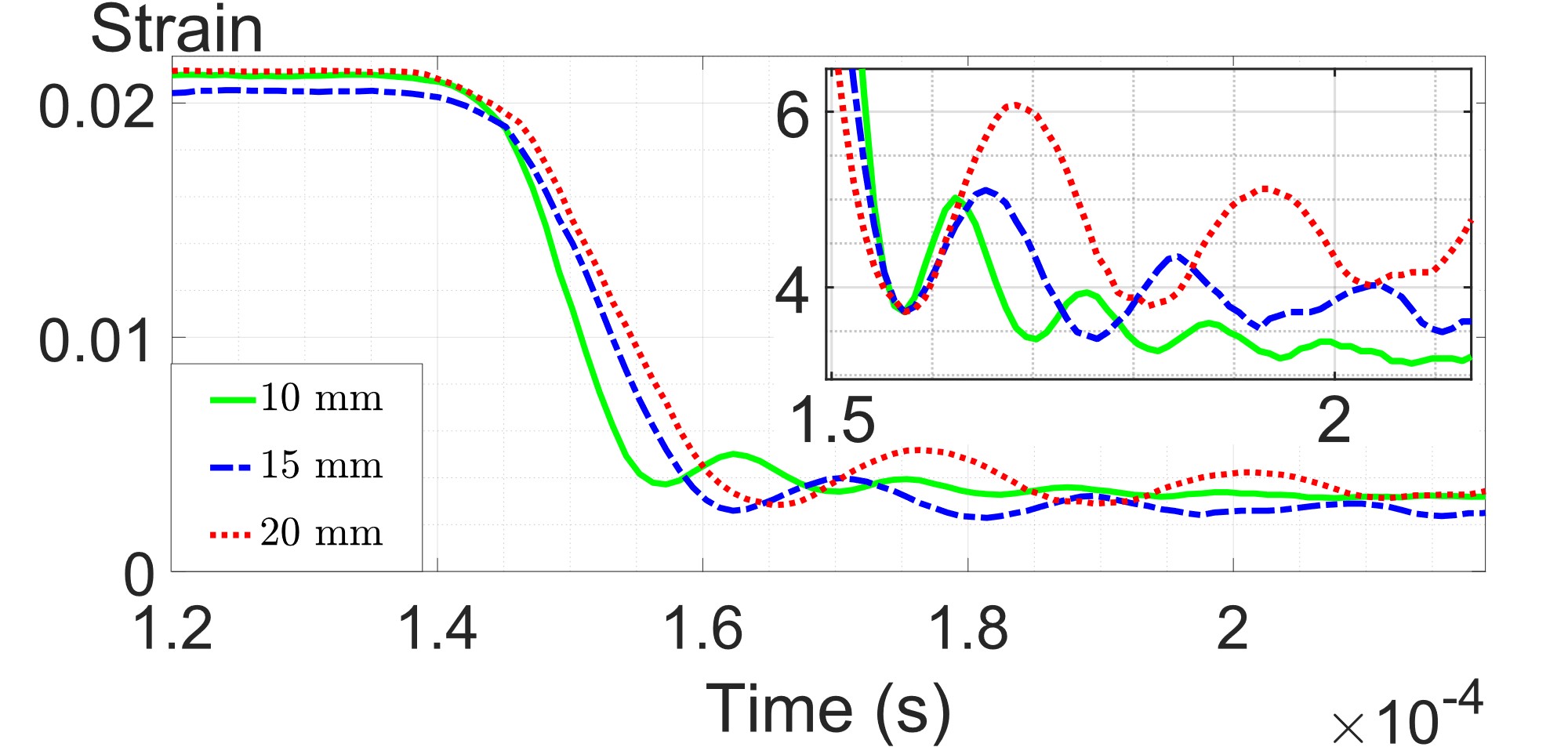}
	\caption{Three experimental strain profiles from bars of different widths, all recorded by CP3 when it was positioned 0.30 m away from the fracture site. In the insert, the blue (dash-dot) and red (dot) curves have been translated in order to provide an easier comparison of the oscillations. }
	\label{fig:fig4}
\end{subfigure}
	
	\begin{subfigure}[t!]{0.5\textwidth}
		\centering
		\includegraphics[width=0.95\textwidth]{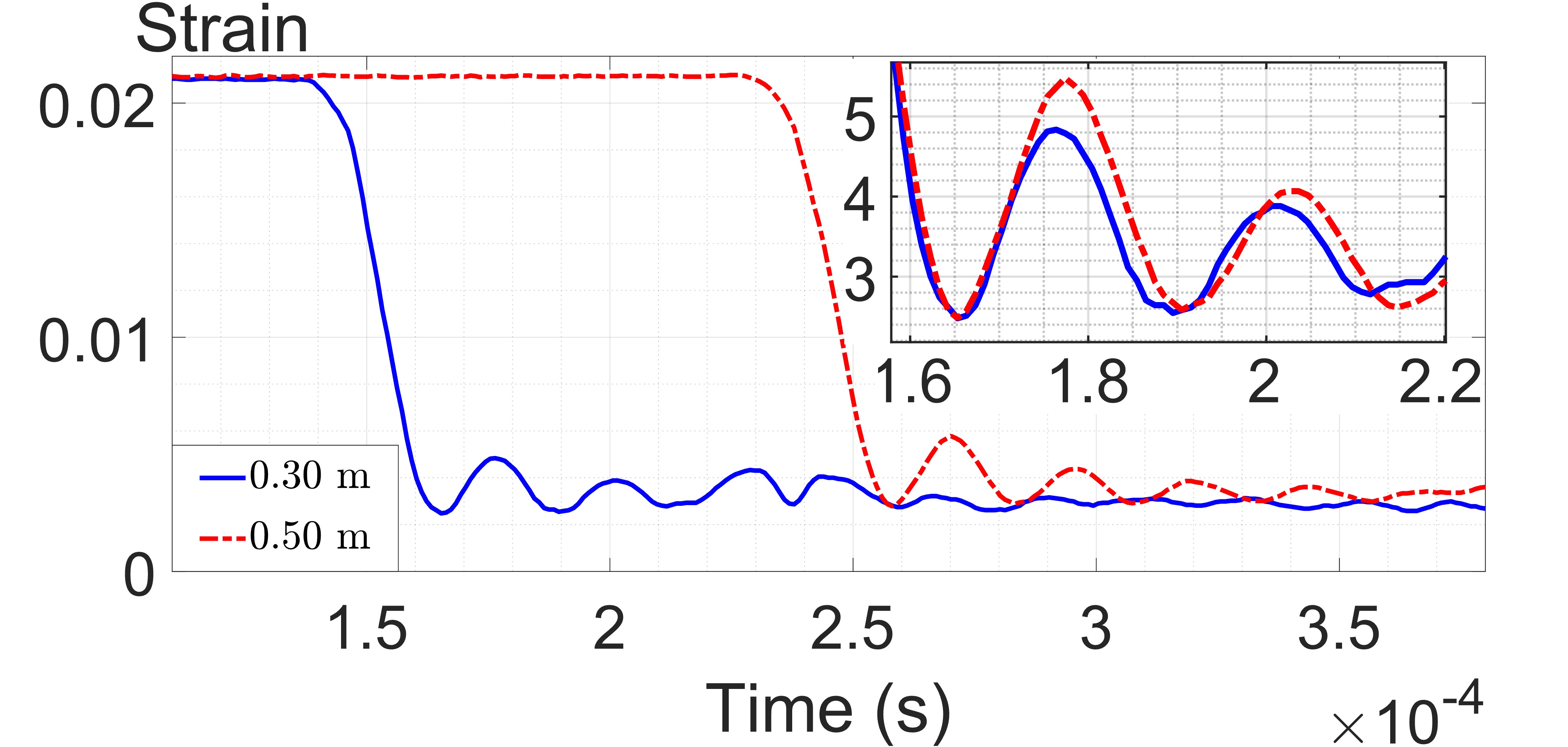}
		\caption{Two experimental strain profiles from 20 mm wide bars at different distances from the fracture site. The profiles are from separate experiments, but the characteristics at fracture were similar ($\dot{e}_5 \sim 1650$ s$^{-1}$).  In the insert, the red (dash dot) curve has been translated in order to provide an easier comparison of the oscillations. }
		\label{fig:fig5}
	\end{subfigure}% \ \
	\caption{A selection of experimental strain profiles illustrating the characteristics of the leading oscillation. }
	\label{fig:3}
\end{figure}

The average speed of the top of the release wave is calculated by taking the value of strain given by $0.95\kappa$ and using the times at which that point is registered by CP1 and CP2. % from the 50 mm and 110 mm profiles is 2140 m/s, and, 2100 m/s from the 50 mm and 300 mm profiles. Thus 
%We observe that the wave slows down as it propagates. 
The details of the profiles in Fig. \ref{fig:2} are given in Table \ref{Tab_1}.

Along with the longitudinal wave, a slower moving shear wave with wave speed $ c_1\sim 1290$ m/s (otherwise referred to as a \emph{flexural} wave) is also generated which 
%decays significantly more 
 shows a much greater variability than the longitudinal wave, as reported in other fracture experiments \cite{Miklowitz, Phillips, Kolsky_1973, K_1976}.In all cases, the amplitude of the flexural wave decreases with the distance from the fracture site.  On the contrary, the longitudinal wave appears to grow and expand  at the distances relevant to our experiments. 

Of particular interest to this paper are the longitudinal oscillations that develop at the bottom of the release wave. These are particularly noticeable at further distances from fracture when there is sufficient separation between the longitudinal wave and the shear wave (see the insert of Fig. \ref{fig:2b}). The leading oscillation is discussed in detail in the next section.

%\begin{figure}
%	\centering
%	\includegraphics[width=0.5\textwidth]{Fig_4}
%	\caption{Three experimental strain profiles from bars of different widths, all recorded by CP3 when it was positioned 0.30 m away from the fracture site. In the insert, the blue (dash-dot) and red (dot) curves have been translated in order to provide an easier comparison of the oscillations. }
%	\label{fig:fig4}
%\end{figure}

\section{Discussion}
From the results of the experiments, it is observed that the features of the leading oscillation that develops at the bottom of the release wave are dependent on numerous parameters. For example, the effect of the strain rate on the oscillation can be seen by comparing measurements with similar pre-strains from bars of the same cross section recorded at the same distance from the fracture site, but different strain rates. Such comparison is shown in Fig. \ref{fig:fig3}. It was observed that the amplitude of the oscillation resulting from the release wave with a steeper slope is greater than that from the shallower slope. Here, amplitude is defined as the vertical distance between the first minimum and first maximum. The duration of the oscillation however, defined as the time between the first and second minima, is approximately the same in both. This suggests %that it is independent of the strain rate. 
there is no strong dependence on strain rate, at least over the fairly limited range considered here.
It was previously observed that the duration at a given distance is also independent of the pre-strain \cite{HRHK}. 

%It is also apparent from 
From Fig. \ref{fig:fig4}, it is apparent that the strain rate of the release wave is shallower in the wider bars. This is not an artefact of the wave propagation as the observation also holds at 0.05 m and 0.11 m from fracture. The decrease in strain rate is due to the tip of the crack travelling a longer distance in the wider bars to cause fracture, thus increasing the duration of the initial unloading, and decreasing the strain rate as a result.

The effect of the width of the waveguide on the duration of the oscillation can be established on comparing measurements taken at the same distance from the fracture site in bars of different cross sections. Measurements at 0.30 m from the fracture site in bars of all three cross sections that were used are shown in Fig. \ref{fig:fig4}. It is observed that the duration of the leading oscillation is longer in the wider bars than in the narrow bars as it is found to be 13 $\mu$s, 19 $\mu$s and 24 $\mu$s in the 10 mm, 15 mm and 20 mm bars respectively. 

It is also noticeable that the %dimension 
width of the bar affects the amplitude of the oscillation on inspection of the strain profiles from 10 mm and 20 mm wide bars. Given their similar pre-strains, had a release wave with the same strain rate of the observed 20 mm profile been generated in a bar of width 10 mm, then it would have a smaller amplitude than the green profile as it would have a shallower release wave. 
%However, the shallower red curve from the 20 mm bar has amplitude approximately double that from the steeper curve in the 10 mm bar. 
However, the red curve from the 20 mm bar has amplitude
approximately double that from the steeper curve in the 10 mm bar, despite the strain rate being significantly lower.

The evolution of the oscillation as it propagates can be established on comparing tests where the third polariscope was positioned at different distances, and where the pre-strains and %strain rates at the early distances was similar. 
strain rates at the shorter distances were similar. 
Such tests are shown in Fig. \ref{fig:fig5}, where the 0.50 m profile from Fig. \ref{fig:2} is shown again, along with a 0.30 m profile with similar fracture characteristics. The amplitude is seen to increase with propagation distance, as too is the duration, %Recall 
which are the two key features of an undular bore. 

%\begin{figure}
%	\centering
%	\includegraphics[width=0.5\textwidth]{Fig_5}
%	\caption{Two experimental strain profiles from 20 mm wide bars at different distances from the fracture site. The profiles are from separate experiments, but the characteristics at fracture were similar ($\dot{e}_5 \sim 1650$ s$^{-1}$).  In the insert, the red (dash dot) curve has been translated in order to provide an easier comparison of the oscillations. }
%	\label{fig:fig5}
%\end{figure}

\subsection{Linear bore}
To model the wave propagation in a uniform rectangular bar, we consider the 
linear Korteweg - de Vries equation with a continuous step initial profile.   This model was obtained by considering the linearisation of the nonlinear Gardner equation \cite{HRHK} near the level of the pre-strain at fracture. While this simple model does not capture all features of the solution, we will show that it allows us to obtain a good semi-quantitative description of the key features at the front of the undular bore. 
The problem is given by
	\begin{equation}
	\begin{cases}
		\displaystyle{e_x + \frac{1}{c_0}e_t - \frac{\delta^2}{2c_0^3} e_{ttt} = 0,} \\
		\displaystyle{e(x_0,t) =  \kappa_t + \frac{\kappa-\kappa_t}{2}\left [ 1 - \text{erf} \left( \frac{t-\eta_1}{2 L} \right) \right ]},
		\label{eqn:linear}
	\end{cases}
\end{equation} 
where $c_0 = \sqrt{E/\rho}$ is the longitudinal wave speed, $\rho$ is density and $\displaystyle{\delta^2 = \frac{(b_1^2 + b_2^2) \nu^2 (1 + 2\nu)}{6(1+\nu)}}$ where $b_1$ and $b_2$ are half the dimensions of the cross section, $\nu$ is Poisson's ratio, $\eta$ is the shift parameter and $L$ is the slope parameter. The initial profile is constructed using the experimental measurement taken closest to the fracture site, which gives $x_0 = 0.05$ m.  The solution to this problem is given analytically by
	\begin{align}
	%V(X,T) = 1 - & \int_\frac{-T}{(3X)^\frac{1}{3}} ^ \infty \text{exp}\Big[ \frac{2(4L^6 + 18 L^2 X (3X)^\frac{1}{3} s)}{108X^2}\Big] \times \nonumber \\ 
	%& \text{Ai}\lb s + \lb \frac{2}{\lb 6X \rb ^4} \rb ^\frac{1}{3} 2L^4 \rb ds,
	%e(x, t) = &\frac{\kappa}{1+\alpha}\Bigg[1 + \alpha -  \int_{b(x, t)} ^ \infty \text{Ai}\left [ s + \lb \frac{2^\frac{1}{3}}{\lb 6x \rb^\frac{4}{3}} \rb  2\tilde{L}^4 \right ]  \nonumber \\ 
	%& \times \text{exp}\Big[ \frac{(2\tilde{L}^6 + 9 \tilde{L}^2 x (3x)^\frac{1}{3} s)}{27x^2}\Big]  ds\Bigg],
	&e (x,t) = \kappa_t + (\kappa - \kappa_t)\Big[ 1-\text{exp}\left(\frac{2L_1^6}{27x^2}\right) \times \nonumber \\ &\int_{b(x, t)}^\infty\text{exp} \left(\frac{sL_1^2}{(3x)^{2/3}}\right)\text{Ai}\left( s + \frac{L_1^4}{(3x)^{4/3}} \right) ds \Big], 
	\label{eqn:my_erf_sol}
\end{align}
%for nondimensional space $X$ and time $T$ 
where  $\displaystyle{b = \left( \frac{2}{3 \delta^2 x} \right) ^{1/3} \left [ x - c_0 \left( t-\eta_1 \right) \right ] }$, $\displaystyle{L_1 = L\left( \frac{2c_0^3}{\delta^2} \right)^{1/3}}$ and Ai is the Airy function \cite{HRHK}  (see also the papers \cite{berry_2019,WT}). From this solution, we can obtain estimates for the development of the slope of the bore front and the leading oscillation by using the experimental measurements taken at 0.05 m and 0.11 m from the fracture site. 

\subsubsection{Amplitude}

%The solution reduces to the integral Airy solution (solution obtained when the initial condition is a step-function \cite{WT}) given by 
%\begin{equation}
%	e (x,t) = \kappa_t + (\kappa - \kappa_t)\Big[ 1- \int_{b(x, t)}^\infty\text{Ai}\left( s \right) ds \Big],
%	\label{eqn:int_airy}
%\end{equation}
%in the limit $L \to 0$, where by the initial condition tends towards a step between the levels of $\kappa$ and $\kappa_t$. For $L \neq 0$, the solution \eqref{eqn:my_erf_sol} also approaches the integral Airy \eqref{eqn:int_airy} in the limit $x \to \infty$. That is regardless of the gradient of the initial condition, i.e. the linear solution forgets its initial slope as it propagates \cite{berry_2019}. 

%The amplitude of the leading oscillation of the integral Airy solution has a maximum value, therefore in the limit $x \to \infty$ the amplitude of the leading oscillation in solution \eqref{eqn:my_erf_sol} will too tend to that maximum value, which in terms of our problem is given by
The maximum amplitude of the leading oscillation of the solution \eqref{eqn:my_erf_sol} obtained in the limit $x \to \infty$ is given by 
	\begin{equation}
	a \approx 0.466(\kappa - \kappa_t).
	\label{eqn:amp}
\end{equation} 
%which is the analogue of the result presented in \cite{berry_2019}.
%Here, $\kappa - \kappa_t$ defines the difference between the values of the pre- and post- strain. For example, from \eqref{eqn:amp} and the data in Table \ref{Tab_1}, the maximum amplitude of the oscillation shown in Fig. \ref{fig:2} is $\sim  8.6  \times 10^{-3}$. At 300 mm from fracture however, the amplitude is $\sim 2.4  \times 10^{-3}$, which is ($\sim 27\%$) of the maximum. The amplitude of the green curve in Fig. \ref{fig:fig4}, also at 300 mm from fracture, is at $15\%$ of its maximum.

%	As this maximum amplitude is only achieved in the limit $x \to \infty$, it is not possible to give a distance from $x_0$ at which this value is reached. A more instructive measure of the growth of amplitude is the distance from $x_0$ at which the value of $a$ reaches some percentage of the maximum \cite{HRHK_2021}. %In particular, the distance at which \eqref{eqn:my_erf_sol} reaches $15\%$ and $27\%$ of the maximum are 
	%is given by
The distance from $x_0$ at which the amplitude reaches $\alpha$\% of the maximum is given in the form
		\begin{equation}
		x_{\alpha} \approx K_\alpha \frac{2c_0^3 L^3}{\delta^2}.
		\label{eqn:amp_perc}
	\end{equation}
%	and
%	\begin{equation}
%		x_{27} \approx 1.2 \frac{2c_0^3 L^3}{\delta^2}  ,
%		\label{eqn:27perc}
%	\end{equation}
%	respectively. 
Particular values of the constant $K_\alpha$ are $K_{15} = 0.75$, $K_{50} = 3.5$, $K_{90} = 56$. Thus for steeper slopes (smaller $L$), given bars of the same cross section (same $\delta$), the amplitude thresholds are reached closer to $x_0$. This is observed in Fig. \ref{fig:fig3} where, at the same distance and in bars of the same geometry, the amplitude of the steeper slope is larger than that from the shallower slope, thus is at a greater percentage of its maximum at a given distance.

\subsubsection{Gradient}

By differentiating the solution \eqref{eqn:my_erf_sol} with respect to $t$, we find the gradient as %of the solution \ref{eqn:my_erf_sol} as
\begin{align}
	e_t(x,t) &= c_0 (\kappa - \kappa_t) \text{exp} \left( \frac{2 L_1^6}{27 x^2}\right) \text{exp} \left( \frac{b L_1^2}{ \left( 3x \right) ^{2/3}} \right) \times \nonumber \\ 
	&\text{Ai} \left( b + \frac{L_1^4}{(3x)^{4/3} }  \right) \left( \frac{2}{3 \delta^2 x} \right)^{1/3}.
	\label{eqn:e_t}
\end{align}
%On evaluating it at $t = \eta_1 + x/c_0$, we have the gradient at a point on the bore front as a function of $x$ as 
%\begin{align*}
%	g_s(x) &=  c_0 (\kappa - \kappa_t)\text{exp} \left( \frac{2 L_1^6}{27 x^2}\right) \times \nonumber \\
%	&\text{Ai}  \left( \frac{ L_1^4}{(3x)^{4/3}} \right)   \left( \frac{2}{3 \delta^2 x} \right)  ^{1/3},
%	\label{eqn:g_s}
%\end{align*}
%where $\displaystyle{L_1 = L\left( \frac{2c_0^3}{\delta^2} \right)^{1/3}}$.
%Again, this qualitatively agrees with the observations in the experiment as it decreases as $x$ increases (the bore gets shallower with propagation distance). 

 \subsubsection{Duration}

From the gradient \eqref{eqn:e_t}, it is clear that stationary points occur only at zeros of the Airy function. On finding the zeros corresponding to the first and second minima, %finding the stationary points corresponding to the first and second minima of the solution \eqref{eqn:my_erf_sol} on differentiation with respect to $t$, 
one can find an estimate of the duration of the first oscillation as
	\begin{equation}
	\hat t (x)  \approx \frac{3.643}{c_0} \left( \delta^2 x \right) ^{1/3}.
	\label{eqn:dur}
\end{equation} 
Immediately, a qualitative agreement to the experimental observations can be seen. Indeed, the duration is not dependent on the strain rate of the release wave, or on $\kappa$ or $\kappa_t$, as was previously observed in the experiments. Rather, it agrees with the observations of the duration being longer at a given distance in wider bars (same $x$, increased $\delta$) as observed in Fig. \ref{fig:fig4}. Also, for the same cross section, the duration gently increases with propagation distance $x$ as seen in Fig. \ref{fig:fig5}.

\section{Estimates}

We now use the profiles that were recorded closest to fracture site  from CP1 and CP2 to obtain the wave speed $c_0$  (i.e. avoiding the need to measure the dynamic Young's modulus $E$ directly), Poisson's ratio $\nu$, $\eta$ and $L$ which are required in order to use formulae (\ref{eqn:amp_perc}), (\ref{eqn:e_t}) and (\ref{eqn:dur}) to provide estimates for the evolution of the bore.  The estimates are then compared to the experimental profile recorded by CP3 to determine their accuracy.

 To illustrate, let us consider the experimental profiles given in Fig. \ref{fig:2}. We take the wave speed $c_0$ as $v_1$ (see Table \ref{Tab_1}) which, along with the shear wave speed $c_1$ which was measured as {1295 m s$^{-1}$}, was then used to calculate $\nu$  through the relation
\begin{equation*}
	c_1 = \frac{c_0}{\sqrt{2(1+\nu)}},
\end{equation*} 
which gives $\nu = 0.34$. Finally, on fitting the error function initial profile in \eqref{eqn:linear} to the portion of the profile recorded from CP1 in Fig. \ref{fig:2a} that was used for calculation of the strain rate, we have $\eta = 29$ $\mu$s and  {\color{black}$L = %2.4 
3.1 \times 10^{-6}$ s.} From $c_0$, the dynamic Young's modulus can be calculated as $E = 4.8$ GPa, where the measured value of the density $\rho = 1060$ kg m$^{-3}$ has been used.

Substituting the values of $\kappa$ and $\kappa_t$ from Table \ref{Tab_1} into formula \eqref{eqn:amp}, we find that the profile in Fig. \ref{fig:2} has a maximum amplitude of $\sim 8.6 \times 10^{-3}$. However, at 0.50 m from the fracture site, the amplitude is at 35\% of its maximum. From (\ref{eqn:amp_perc}), the distance from $x_0$ at which this amplitude percentage will be obtained is $\sim 0.41$ m, where $K_{35} = 1.8$ has been used. Hence, on including the distance of 0.05 m from the fracture site to $x_0$, this estimate for the growth of the amplitude of the oscillation is in close agreement to the experimental observation. 

Also note that the maximum amplitude of the blue (solid) profile in Fig. \ref{fig:fig5} is also $\sim 8.6 \times 10^{-3}$, whilst at 0.30 m, the amplitude is at 27\% of this maximum. On using the same parameter values as in the previous estimate, we obtain from formula (\ref{eqn:amp_perc}) that $x_{27}\sim 0.29$ m, where $K_{27} = 1.3$ has been used. 

 The amplitude of the green (solid) curve in Fig. \ref{fig:fig4}, also at 0.30 m from fracture, is at $15\%$ of its maximum.
Using the same values of $c_0$ and $\nu$ as %previous, 
before, but with the newly fitted value of $L = 2.3 \times 10^{-6}$ s$^{-1}$ obtained by using the relevant strain profile from CP1, %eta = 3.004 e-05
we have $x_{15} \approx 0.28$ m ($K_{15} = 0.75$). All estimates for the growth of amplitude are in close agreement to the experiment, and one and the same values of $c_0$ and $\nu$ have been used.

%%%%%%%%%% grdient
To estimate the gradient of the bore front at some distance $x_1$ from the fracture site, we again turn to the profiles recorded by CP1 and CP2. On using the value of $0.5\left(\kappa + \kappa_t\right)$ and $c_0$, the time $t_1$ at which this point will arrive at the distance in question can be calculated. 
%We first find the midpoint of the region used to calculate the strain rate (and also to fit the parameters $\eta$ and $L$), and use $c_0$ to obtain an approximate time $t_1$ at which this point will arrive at the distance in question.

On substituting the experimental values from Table \ref{Tab_1} %and the fitted value of $L$ to the profile shown in Fig. \ref{fig:2} 
with the corresponding fitted value of the slope parameter $L = %2.4 
3.1 \times 10^{-6}$ s and evaluating \eqref{eqn:e_t} at $x_1 = 0.45$ m and $t_1 = 24.1$ $\mu$s, %to correspond region used to calculate the strain rate of the experimental data, 
we find that $e_t(x,t) \sim 1100$ s$^{-1}$. This is in excellent agreement with the corresponding strain rate at 0.50 m as given in Table \ref{Tab_1} for the relevant experimental profile.

 Equivalent numerical treatment of the strain profiles recorded by CP1 and CP2 for the green profile in Fig. \ref{fig:fig4} (from a 10 mm wide bar) gave $\eta = 30$ $\mu$s and $L = 2.3\times 10^{-6}$ s$^{-1}$. Then, on using the same procedure and value of $c_0$ as used in the previous estimate for a 20 mm wide bar, we find an estimate of the strain rate at 0.30 m from fracture as 1680 s$^{-1}$, which is again close to the experimental value of 1610 s$^{-1}$.

%%%%%%%%%Duration
From Fig. \ref{fig:fig5}, the duration at 0.30 m and 0.50 m is %$2.42 \times 10^{-5}$
$24$ $\mu$s and %$2.52 \times 10^{-5}$ 
$25$ $\mu$s respectively. %, giving a growth over the 200 mm propagation distance between the two profiles of $1 \times 10^{-6}$ s. 
From \eqref{eqn:dur}, we estimate that the duration of the oscillation at 0.30 m and 0.50 m from fracture are $\hat{t}_{30} = 15.6$ $\mu$s and  $\hat{t}_{50} = 18.5$ $\mu$s respectively. %, where $c_0 = 2110$ m/s was obtained from the 50 mm and 110 mm strain profiles. 
These are slight under estimates compared to the experimental data, which is also observed when estimates of other tests are compared. The estimate of the growth in the duration over this distance is an over estimate.

We note that in this simple linearised elastic model discussed here, we have only the leading order dispersive term accounting for the dispersion due to the geometry of the bar. Modelling of the full problem should contain nonlinear terms and higher order dispersive terms,  which have been found to be important \cite{HRHK, GBK}. Viscoelasticity is also omitted here which too introduces additional dispersive  and dissipative effects \cite{H,HRHK}.

%%%%%% amplitude from another curve 
 %Therefore if the initial profile constructed around the 50 mm experimental profile is take as the condition at the fracture site ($x = 0$), the estimate of the grown of amplitude is very close to what is observed in the experiment. 

\section{Conclusion}
We have generated an undular bore in uniform rectangular PMMA bars of constant cross section by induced tensile fracture. The waves were recorded with high speed multi-point photoelasticity. Our observations are that the evolution of the bore is dependent on the characteristics of fracture and the geometry of the rectangular cross section. From the analytical solution to the linear Korteweg -  de Vries equation with a smooth step initial profile, simple formulae for the key characteristics of the leading oscillation of the bore have been derived. 

On using the strain rate of the lower part of the strain profile recorded 0.05 m from the fracture site, and the wave speed of the top of the release wave between 0.05 m and 0.11 m from the fracture site, estimates for the development of the bore at later distances have been calculated and compared directly to experiments. The estimates for the growth of the amplitude and decrease of the slope are in excellent agreement with experimental observations. To make these estimates, which were accurate for bars of different widths, one and the same value of the  dynamic Young's modulus was used. A qualitative agreement has been established for the growth of the duration and its dependence on the geometry of the cross section.  Higher-order dispersive terms need to be accounted for in order to improve the estimates of the duration \cite{HRHK}.

We anticipate that the formulae derived would be particularly well suited to giving useful estimates for the evolution of the wave in linearly elastic materials such as steel. We expect such waves to be present in the signals generated 
%by earthquakes, fracking and other similar events in situations 
 in various natural and industrial settings involving transverse fracture of the pre-strained waveguides. 

\vspace{1cm}

%\nocite{*}
\bibliography{aipsamp}% Produces the bibliography via BibTeX.

\end{document}